\documentclass[aps,prl,twocolumn,showpacs,preprintnumbers,amsmath,amssymb,superscriptaddress,longbibliography,nofootinbib]{revtex4-2}
\usepackage{hyperref}
\hypersetup{colorlinks=true, citecolor=blue, urlcolor=blue, linkcolor=blue,urlcolor=cyan}
\usepackage{color}
\usepackage{graphicx}% Include figure files
\usepackage{dcolumn}% Align table columns on decimal point
\usepackage{braket}
\usepackage{tikz}
\usepackage{physics}

\newcommand{\grethree}{\mathrm{I}\hspace{-1.2pt}\mathrm{I}\hspace{-1.2pt}\mathrm{I}}
\usepackage{braket}
\newcommand{\be}{\begin{equation}}
\newcommand{\ba}{\begin{eqnarray}}
\newcommand{\ea}{\end{eqnarray}}
\newcommand{\ee}{\end{equation}}

\newcommand{\bea}{\begin{eqnarray}}
\newcommand{\eea}{\end{eqnarray}}
\newcommand{\bes}{\begin{equation*}}
\newcommand{\beas}{\begin{eqnarray*}}
\newcommand{\eeas}{\end{eqnarray*}}
\newcommand{\bas}{\begin{array*}}
\newcommand{\eas}{\end{array*}}
\newcommand{\ees}{\end{equation*}}
\newcommand{\nn}{\nonumber}

\newcommand{\ep}{\epsilon}

\newcommand{\dcN}[0]{\mathcal{M}_{\mathrm{dc}}^{\mathbb{Z}_N}}
\newcommand{\dc}[0]{\mathcal{M}_{\mathrm{dc}}}

\makeatother
\usepackage{comment}
\usepackage{babel}

\allowdisplaybreaks

\begin{document}

\title{Entanglement Entropy via Double Cone Regularization}
\preprint{YITP-23-170}

\author{Taishi Kawamoto}\email{taishi.kawamoto@yukawa.kyoto-u.ac.jp}
\affiliation{\it Center for Gravitational Physics and Quantum Information, Yukawa Institute for Theoretical Physics, Kyoto University, Kitashirakawa Oiwakecho, Sakyo-ku, Kyoto 606-8502, Japan}

\author{Yu-ki Suzuki}\email{yu-ki.suzuki@yukawa.kyoto-u.ac.jp}
\affiliation{\it Center for Gravitational Physics and Quantum Information, Yukawa Institute for Theoretical Physics, Kyoto University, Kitashirakawa Oiwakecho, Sakyo-ku, Kyoto 606-8502, Japan}

%%%%%%%%%%%%%%%%%%%%%%%%%%%%%%%%%%%%%%%
%%%%%%%%%%%%%%%%%%%%%%%%%%%%%%%%%%%%%%%
%\date{\today}

\begin{abstract}
%We find useful way of computing the entanglement Renyi entropy applicable for any quantum field theory. 
This paper proposes an alternative regularization method for handling the ultraviolet behavior of entanglement entropy. Utilizing an  $i\epsilon$ prescription in the  Euclidean double cone geometry, it accurately reproduces the universal behavior of entanglement entropy. The method is demonstrated in the free boson theory in arbitrary dimensions and two-dimensional conformal field theories. 
%Notably, in even dimensions, the $i\epsilon$ prescription results in an imaginary constant term, implying an imaginary boundary entropy in two dimensions. 
The findings highlight the effectiveness of the $i\epsilon$ regularization method in addressing ultraviolet issues in quantum field theory and gravity, suggesting potential applications to other calculable quantities.
\end{abstract}
\maketitle

%%%%%%%%%%%%%%%%%%%%%%%%%%%%%%%%%%%%%%%
%\section{Introduction}
%%%%%%%%%%%%%%%%%%%%%%%%%%%%%%%%%%%%%%%

\noindent \emph{1.Introduction and Summary.}
 In the realm of quantum field theory, particularly within the framework of Feynman's path integral \cite{Feynman:1948ur,Feynman:1949zx}, the historical exploration of the $i\ep$ prescription for the propagator has yielded valuable insights. This technique, also employed to formulate the wave-functional of the vacuum state of the universe \cite{Hartle:1983ai}, has more recently found in the regularization of the tip of a double cone geometry, as observed in the spectral form factor analysis \cite{Saad:2018bqo,Witten:2021nzp,Chen:2023hra}. This paper delves into the utilization of the $i\ep$ prescription in the Euclidean version of the double cone geometry, offering an alternative derivation of the universal behavior of the entanglement entropy.

The entanglement entropy, introduced in quantum information theories and also in quantum field theories \cite{Holzhey:1994we}, measures the entanglement between subsystems. A breakthrough in its calculation emerged with the introduction of twist operators \cite{Calabrese:2004eu}, providing a pathway to derive the universal logarithmic behavior of entanglement entropy in two-dimensional conformal field theory (CFT). Significantly, entanglement entropy diverges in quantum field theories due to the infinite degrees of freedom inherent in the vacuum state, necessitating the introduction of an ultra-violet (UV) cut-off for its quantification.

For a $d$-dimensional quantum field theory, the entanglement entropy of low-energy states is anticipated to follow the form from the holographic calculation \cite{Ryu:2006bv,Ryu:2006ef}:
\begin{align}\label{eq:EE expansion}
    S_{\mathcal{A}}= \frac{C_{d-2}}{\ep^{d-2}}+\cdots+
    \begin{cases}
       C_0\log{\frac{\xi}{\ep}}+\cdots,& \text{for}\,\,\text{even $d$}\\
      (-1)^{\frac{d-1}{2}}F+\cdots. &\text{for}\,\, \text{odd $d$}
    \end{cases}
\end{align}
,where $\xi$ is a typical correlation length of the theory. The initial term illustrates the area law of entanglement entropy, prevalent at low energies. Remarkably, the logarithmic term in even dimensions is proven to be universal. Also in odd dimensions the constant term exhibits universality. The term "universal" is employed to convey its independence of the UV regularization schemes.

This paper introduces an approach to extract the universal term of entanglement entropy through the Euclidean double cone calculations. By employing the $i\ep$ prescription in the metric near the $\mathbb{Z}_N$ orbifold singularity (where $1/N$ represents the number of replica sheets), we establish the efficacy of this regularization method as an UV cut-off to entanglement entropy. This assertion aligns with \cite{Ohmori:2014eia}, emphasizing the necessity of imposing boundary conditions around the entangling surface (the end points of the subsystem) to define entanglement entropy properly, given a  division of the Hilbert space into subsystems. A distinctive advantage of our method lies in the compatibility with the Heat kernel method, even with the introduction of a regulator. In contrast, approaches involving a cut-off in the geometry, such as the brick wall case \cite{tHooft:1984kcu}, face challenges in solving analytically via the method of images. Moreover, what kind of physically relevant boundary conditions should be favored remains unclear. 
On the other hand, in our computation, we do not care about these boundary problems since we have a complex "wormhole" rather than solid boundaries. To solve the wave equation or diffusion equation for the heat kernel, we are demanding, for example, fall-off conditions for ${r\to -\infty}$ and ${r\to \infty}$ asymptotic in the Minkowski space we see later.
To circumvent these difficulties, we extract the half cone contribution by halving the result in evaluating the partition function. 

We demonstrated in the free scalar theory on a flat space-time in any dimensions and arbitrary two-dimensional CFTs. Despite successfully reproducing universal terms, our approach comes with a trade-off. Specifically, in our calculation of the entanglement entropy for the free boson theory in even dimensions, we encounter an imaginary constant term. Similarly, in odd-dimensional cases, the non-universal part becomes purely imaginary. This outcome aligns with the inherent complexity of the Schwinger parameter in partition function evaluation and the non-Hermitian nature of the modular Hamiltonian for the sub-algebra, as noted in \cite{Chen:2023hra}\footnote{We may have some options to extract real value from these in general complex quantities, such as taking absolute values or real or imaginary parts, though we need to check whether they are proper measure of entanglement.}. While our double cone regularization method yields the same universal terms compared to other methods such as the momentum cut-off or the lattice regularization, it operates through a totally different manner. In particular, the spectrum of modular Hamiltonian is drastically changed if we turn on the regularization parameter. It is important to stress that the modified modular Hamiltonian have complex spectrum and quasi normal modes, which capture the universal anomaly terms though the spectrum itself does not depend on regularization parameter in an explicit way. Understanding how to extract physically meaningful quantities amid the renormalization of complex parameters represents an interesting avenue for future exploration. For the scope of this paper, our focus remains on the universal terms.

It is pertinent to draw attention to the parallelism with pseudo entropy \cite{Nakata:2020luh,Mollabashi:2020yie}.
%time-like entanglement entropy  \cite{Doi:2022iyj}, where a real universal logarithmic term and a purely imaginary constant term, proportional to the central charge, are identified. Our situation mirrors this scenario closely. Notably, our calculations, particularly in  the two-dimensional free boson theory, align with that of time-like entanglement entropy.
We anticipate that the specific details of the complex contour deformation will yield distinct signs and values for the constant terms\footnote{In this case, we consider a final state post-selection, which makes physics
non-Hermitian.}. Pursuing this avenue, we find it intriguing to investigate its applications to the holographic entanglement entropy \cite{Ryu:2006bv,Ryu:2006ef} and also to de Sitter holography \cite{Strominger:2001pn}.  The utility of the $i\ep$ prescription in the metric extends to various situations where analytical solutions are favored while introducing a cut-off.
\vspace{1 pt}

\noindent \emph{2.Entanglement entropy in free scalar fields.}
In the subsequent section, we revisit the orbifold method employed in deriving entanglement entropy as outlined in \cite{Nishioka:2006gr}. Suppose we would like to compute the entanglement entropy of the half-space $\mathcal{A}\; ,x_1>0$ in the Minkowski space $\mathbb{R}^{1,d-1}$,
    \begin{equation}\label{eq:Mink}
        ds^2 = -d{x_0}^2 + dx_1^2 + dx_{\mathbb{R}^{d-2}}^2. 
    \end{equation}  To use the replica method, we do the Wick rotation $x_0\to i\tau$. For ease of representation, we amalgamate $\tau$ and $x_1$ into a single complex plane $\mathbb{C}$. Employing the orbifold method, we calculate the $n$-th R\'{e}nyi entropy for the half-infinite region.

Utilizing the orbifold method, the $n$-th R\'{e}nyi entropy for a semi-infinite subsystem is derived as follows:
\be
S_\mathcal{A}^{(n)}=\frac{1}{1-n}\left[{Z\qty[\mathbb{C}/\mathbb{Z}_N\times\mathbb{R}^{d-2}]}-\frac{1}{N}{Z\qty[\mathbb{C}\times\mathbb{R}^{d-2}]}\right]_{N=\frac{1}{n}},\label{partition}
\ee
where  the action of $\mathbb{Z}_N$ follows
\be
 X=x_1+i\tau\rightarrow Xe^{\frac{2\pi i}{N}}.
\ee
and $Z[\mathcal{M}]$ denotes the log of the partition function of QFT (free scalar, for example) on the manifold $\mathcal{M}$. In this context, the partition function corresponds to the one at the first quantization. Employing the heat kernel method, as extensively reviewed in \cite{Vassilevich:2003xt}, especially we use the expression for the heat kernel in flat space $\mathbb{R}^d$;
  \begin{equation}
      K_{\mathbb{R}^d}(x,x';t)= \frac{1}{(4\pi t)^{\frac{d}{2}}}e^{-\frac{r^2}{4 t}-tm^2}, \quad r=\abs{x-x'},
  \end{equation}
  where $t$ is the Schwinger parameter. 
Also the expression for the heat kernel for the orbifold is obtained via method of images as
\begin{equation}
    \begin{split}
&K_{\mathbb{C}/\mathbb{Z}_N\times\mathbb{R}^8}(r,\theta,x_i;r',\theta',x'_i;t) \\&= \frac{1}{N}\sum_{k=0}^{N-1} K_{\mathbb{C}\times\mathbb{R}^{d-2}}\qty(r,\theta,x_i;r',\theta'-\frac{2\pi k}{N},x'_i;t) .
    \end{split}
\end{equation}
From this we can evalaute the Renyi n-th entropies as 
\begin{align}
   S_\mathcal{A}^{(n)}&=\int_{\ep^2}^\infty\frac{dt}{2t}\int d^dx\sqrt{g}\left(K_{\mathbb{C}/\mathbb{Z}_N\times \mathbb{R}^{d-2}}(x,x;t )\right.\nn\\
   &\left.-K_{\mathbb{C}\times \mathbb{R}^{d-2}}(x,x;t)\right)|_{N=1/n}\nn\\
    &=\frac{(n+1)\pi V_{d-2}}{6n} \int_{\ep^2}^\infty \frac{dt}{(4\pi t)^{\frac{d}{2}}} e^{-m^2 t}\nn\\
    &=\frac{(n+1) }{6n }\frac{\pi}{(4\pi)^{\frac{d}{2}}}\frac{V_{d-2}}{\ep^{d-2}} E_{\frac{d}{2}}\qty(\ep^2 m^2),\label{int}
\end{align}
where $E_\nu(z)$ is the $\nu$-th order exponential integral and we introduce a cut-off $\ep$.
We can expand the $S_{\mathcal{A}}^{(n)}$ by $\ep^2$. The universal term is given by the constant term and logarithmic term for odd and even $d$, respectively 
\begin{equation}\label{eq:momentum RE universal}
   S_{\mathcal{A}}^{(n)}
   \sim 
   \begin{cases}
       
   \frac{(n+1) }{6n }\frac{\pi}{(4\pi)^{\frac{d}{2}}}\Gamma\qty(\frac{2-d}{2})\cdot V_{d-2}\;m^{d-2},&  \text{$d$ is odd}\\
       \frac{(n+1) }{6n }\frac{2\pi}{(4\pi)^{\frac{d}{2}}}\frac{(-1)^{\frac{d}{2}-1}}{\qty(\frac{d}{2}-1)!}\cdot V_{d-2}\;m^{d-2}\log\qty(\frac{1}{m\ep}), &  \text{$d$ is even.}
   \end{cases}\nn
\end{equation}

%\subsection{Double cone calculation
Above we manually introduced the cut-off scale in (\ref{int}) through dimensional analysis. A natural question to ask is whether we can interpret this $\epsilon$ as a geometric cut-off. The $i\epsilon$ prescription provides a lucid understanding of this. Below, we demonstrate that this prescription effectively resolves the orbifold singularity.
\par 
Again we would like to compute the entanglement entropy of the half-space $\mathcal{A}$ of the Minkowski space \eqref{eq:Mink}. 
    As another coordinate for this entire Minkowski spacetime, we will have Rindler coordinates,
    \begin{equation}\label{eq:Rindler_coordinate}
        ds^2 = dr^2 - r^2 dT_{\mathrm{Rindler}}^2 + dx_{\mathbb{R}^{d-2}}^2,
    \end{equation}
    where the Rindler radial coordinate $r$ runs $-\infty < r < \infty$ and we denote $T_{\mathrm{Rindler}}$ as a time coordinate of the Rindler spacetime. The region $r>0$ corresponds to the right Rindler wedge and $r<0$ corresponds to the left Rindler wedge.  Then, we consider the QFT on this Minkowski spacetime or, equivalently, the Rindler spacetime. However, similar to the QFT on a black hole background, there is a UV divergence from the horizon. To regularize this UV divergence from the horizon, we consider the complex deformation,
    \begin{equation}
        r \in \mathbb{R} \to r = \tilde{r} - i\epsilon,\; \epsilon > 0,\; \tilde{r} \in \mathbb{R}.
    \end{equation}
    Also, to perform the replica trick or the orbifold trick, we consider the Wick rotation of the Rindler time $T_{\mathrm{Rindler}}$ as $T_{\mathrm{Rindler}} \to i\theta$. Then, by rewriting $\tilde{r}$ as $r$, we obtain the metric 
    \begin{equation}\label{eq:our_double_cone}
        ds^2 = dr^2 + (r - i\epsilon)^2 d\theta^2 + ds^2_{\mathbb{R}^{d-2}},\; r\in (-\infty,\infty).    \end{equation}
        We denote this manifold as $\mathcal{M}_{\mathrm{dc}}$. This kind of the metric is introduced in \cite{Saad:2018bqo, Witten:2021nzp, Chen:2023hra}. 
     Also, to compute the $n$-th Renyi entropy, we consider the $\mathbb{Z}_N$ orbifold at $r=0$ (before the complex deformation), which is realized as an identification $\theta \sim \theta + \frac{2\pi}{N}$. Then, we have the geometry, 
     \begin{equation}\label{eq:our_double_cone_ZN}
        ds^2 = dr^2 + (r - i\epsilon)^2 d\theta^2 + ds^2_{\mathbb{R}^{d-2}},\; \theta \sim \theta + \frac{2\pi}{N}.
    \end{equation}
     We denote this manifold as $\mathcal{M}_{\mathrm{dc}}^{\mathbb{Z}_N}$. The heat kernel on $\dc$ and $\dcN$, $K_{\dc}(x,x';t)$ and $K_{\dcN}(x,x';t)$ are obtained by analytic continuation from $K_{\mathbb{R^d}}(x,x';t)$ and $K_{\mathbb{C}/\mathbb{Z}_N\times\mathbb{R}^{d-2}}(x,x';t)$ respectively. 
 We claim that the metric, utilizing the $i\epsilon$ prescription, not only satisfies the Einstein equation but also offers a more flexible method for regulating the UV divergence though initially we do not necessarily introduce an imaginary regulator. \par
We now delve into the details of calculating the entanglement entropy. To make our procedure of scalar field theory clear, let us provide a more detailed exposition of the paper's content. We stated that the $n$-th Renyi entropy is given by a half of the partition function on the orbifold \eqref{eq:our_double_cone_ZN}, as follows:
\begin{equation}\label{eq:def_Renyi}
    S_{\mathcal{A}}^{(n)}= \frac{1}{2(1-n)}\qty[{Z[\dcN]}-\frac{1}{N}{Z[\dc]}]_{N=\frac{1}{n}}.
\end{equation}
 Since our subsystem $\mathcal{A}$ resides in $r>0$, it may seem tricky, or one might feel that we are "extending" spacetime from $r>0$ to $-\infty <r<\infty$ and deforming it in a complex manner. However, in our double cone regularization, we expect \eqref{eq:def_Renyi} computes the Renyi entropy. One possible explanation is as follows.
Suppose we have a entanglement state $\psi$ in a total space and consider the reduced density matrix of the state $\psi$ in $r>0$ and $r<0$ denoted by $\rho_{r>0}$ and $\rho_{r<0}$, respectively. 
Then we prepare a operator $\rho$ such that
\begin{equation}\label{eq:state}
    \rho = \qty(\rho_{r>0})\otimes (\rho_{r<0}).
\end{equation}
Note that $\Tr_{r>0}{\qty[(\rho_{r>0})^n]}$ and $\Tr_{r<0}{\qty[(\rho_{r<0})^n]}$ 
 will have a UV divergence. Geometrically, the $\rho$ will prepare two Euclidean cone geometries. Its partition function is given by:
\begin{equation}
    \Tr\qty[\rho^n] = \Tr_{r>0}\qty[(\rho_{r>0})^n]\Tr_{r<0}\qty[(\rho_{r<0})^n].
\end{equation}
If we assume symmetry between $\rho_{r>0}$ and $\rho_{r<0}$ (as in the thermo field double case),  the left wedge $r<0$ and the right wedge $r>0$ are equivalent,  we have
\begin{equation}
\Tr_{r>0}\qty[(\rho_{r>0})^n]=\Tr_{r<0}\qty[(\rho_{r<0})^n]
\end{equation}
and thus
\begin{equation}
    \Tr\qty[\rho^n] = \Tr_{r>0}\qty[\qty(\rho_{r>0})^n]^2.
\end{equation}
We insist that this toy explanation is just for showing that in simple usual system the Renyi entropy computation \eqref{eq:def_Renyi} matches the usual computation. We insist that this toy explanation is just for showing that in simple usual system the Renyi entropy computation \eqref{eq:def_Renyi} matches the usual computation. We do not completely expect that this is  true since the factorization of the Hilbert space may not hold in our double cone regularization. This is an interesting direction and related to the long-standing puzzle of the factorization problem but we do not delve in this problem here.  Even if there may not exist  $\rho_{r>0}$ and $\rho_{r<0}$, we think that we compute $\Tr{\rho^n}$ in our double cone regularization and we expect the right hand side of \eqref{eq:def_Renyi} computes the Renyi entropy for the subsystem $\mathcal{A}$. This point is verified to some extend since we can derive the universal term of the Renyi entropy in this regularization. 
Before the deformation, the  partition function has UV divergences. Thus, we need to regularize them in some way. Conceptually our double cone regularization of $\rho$ connects the two Euclidean path integrals $\Tr_{r>0}{[(\rho_{r>0})^n]}$ and $\Tr_{r<0}{[(\rho_{r<0})^n]}$ by a complex "wormhole". 
This construction is essentially similar to the discussion
\cite{Anegawa:2021osi,Anegawa:2022pce}, where they discuss the wormhole in the real Euclid spacetime rather than our complex wormhole.

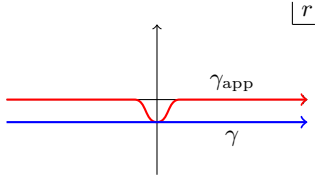
\begin{figure}[h]
    \centering
    \begin{tikzpicture}
\draw[->](-2,0)--(2,0);
\draw[->](0,-1)--(0,1);
\draw[red,thick,->](-2,-0)--(-0.3,-0)to[out=0,in=180](0,-0.3)to[out=0,in=180](0.3,0)--(2,0);
\draw[blue,thick,->](-2,-0.3)--(2,-0.3);
\draw(1.8,1)--(2.1,1);
\draw(1.8,1)--(1.8,1.3);
\draw(2,1)node[above]{$r$};
\draw(1,0)node[above]{$\gamma_{\mathrm{app}}$};
\draw(1,-0.3)node[below]{$\gamma$};
\end{tikzpicture}
    \caption{The contour $\gamma$ for our double cone regularization. We can also consider another complex contour $\gamma_{\mathrm{app}}$ which is used in the Appendix \ref{appA}. }
    \label{fig:r-contour}
\end{figure}

Specifically, we apply the orbifold method and conduct a volume integral of the heat kernel. As a result, we obtain:
\begin{equation}
\begin{split}
    &Z\qty[\dcN]-\frac{1}{N}Z\qty[\dc]\nn\\
&=\int_0^\infty\frac{dt}{2t}\int d^dx\sqrt{g}\left(K_{\dcN}(x,x;t)-K_{\dc}(x,x;t)\right)\nn\\
    &=\frac{2\pi V_{d-2}}{N}\sum_{k=1}^{N-1}\int_{\Gamma}\frac{dt}{2t}\frac{1}{(4\pi t)^{\frac{d}{2}}}\int_\gamma dr\sqrt{r^2} e^{-tm^2-\frac{r^2}{t}\sin^2{\frac{\pi k}{N}}},\nn
\end{split}
\end{equation}
where $g$ is the determinant of the metric and $k=0$ contribution of $Z[\dcN]$ is canceled by $Z[\dc]$. This cancellation ensures us that there is no IR-divergence from $r$-direction. We choose a contour of radius $r$, denoted as $\gamma$, in the complex plane of Fig.\ref{fig:r-contour}. The integration with respect to $r$ can be carried out as follows: let us deform $\gamma$ to $r-i\ep$, where $-\infty<r<\infty$, to evaluate the integral. It is crucial to handle the branch of $\sqrt{r^2}$. To obtain a non-zero result, we select the negative branch for $\mathrm{Re}\qty[r]<0$ part. Since the integration on the deformed contour is $\mathbb{Z}_2$ invariant ($\Re{r}\to-\Re{r}$), we consider only the part $\mathrm{Re}\qty[r]>0$. Consequently, we obtain:
\begin{equation}\label{eq:r-integral}
\begin{split}
    &Z\qty[\dcN]-\frac{1}{N}Z\qty[\dc]\nn\\
&= \frac{\pi V_{d-2}}{(4\pi)^{\frac{d}{2}} N} \sum_{k=1}^{N-1}\frac{1}{\sin^2{\qty(\frac{k\pi}{N})}} \int_\Gamma dt\cdot t^{-\frac{d}{2}} e^{-tm^2 +\frac{\ep^2}{t}\sin^2{\qty(\frac{k\pi}{N})}}.
\end{split}
\end{equation}
%%%%%%%%%%%%%%%%%%%%%%%%%%%%%%%%%%%%%

%%%%%%%%%%%%%%%%%%%%%%%%%%%%%%%%
Suppose we take $\Gamma=[0,\infty)$ on the real axis. Then we see the singular divergence around $t\to 0+$ from the exponential $\exp{\frac{\epsilon^2}{t}\sin{\frac{ k \pi}{N}}}$ since the sign in the exponential is positive. To obtain the finite result without introducing another cutoff, Therefore, we must consider $\Gamma$ in the complex plane. This implies that we need to involve complex Schwinger parameters when dealing with the complex metric space-time \cite{Witten:2013pra}. Especially, we require that in the contour we approach $t=0$ from $\arg{t} \sim \pi$. Also, for simplicity, we choose on $\Gamma$ the parameter to approach $\Re[t]\to \infty$ and $\Im[t]\to 0$.   
 \par
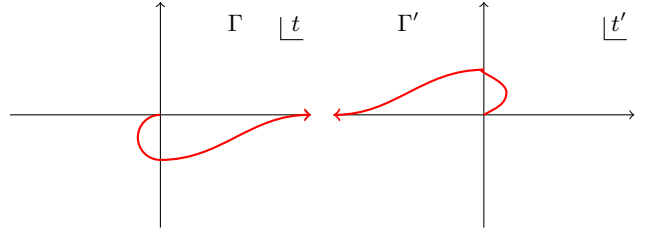
\begin{figure}[h]
    \centering
        \begin{tikzpicture}
\draw[->](-2,0)--(2,0);
\draw[->](0,-1.5)--(0,1.5);
\draw[red,thick,->](0,0)to[out=180,in=90](-0.3,-0.3)to[out=-90,in=180](0,-0.6)to[out=0,in=-180](2,0);
\draw(1.9,1)--(1.6,1)--(1.6,1.3);
\draw(1.8,1)node[above]{$t$};
\draw(1,1)node[above]{$\Gamma$};
\draw[->](2.3,0)--(6.3,0);
\draw[->](4.3,-1.5)--(4.3,1.5);
\draw[red,thick,->](4.3,0)to[out=30,in=-90](4.6,0.3)to[out=90,in=180](4.3,0.6)to[out=180,in=0](2.3,0);
\draw(1.9+4.3,1)--(1.6+4.3,1)--(1.6+4.3,1.3);
\draw(1.8+4.3,1)node[above]{$t'$};
\draw(-1+4.3,1)node[above]{$\Gamma'$};
\end{tikzpicture}
\caption{(Left):The contour for the Heat kernel. (Right):The contour for the Hankel function.}
\label{fig:tcontour}
\end{figure}

 
  As a solid example, let us choose a contour which naturally appears from the relation of the heat kernel and the Green function. In determining the integral contour for the complex Schwinger parameter, we remind a fundamental identity as described in \cite{Mann:1996ze},
 \begin{equation}
     \int_{\Gamma} dt\; K_{\dc}(x,x';t)= G_{\dc}(x,x'),
 \end{equation}
 where $G_{\mathcal{M}}(x,x')$ is a Green's function, satisfying
 \begin{equation}
     (\Box_x-m^2)G_{\mathbb{\mathcal{M}}}(x,x')= \frac{1}{\sqrt{g(x)}}\delta(x-x').
 \end{equation}
 We can calculate the Green's function and the heat kernel separately. From this identity, we can specify the appropriate contour for the Schwinger parameter. 
  The Green's function for $\mathbb{R}^d$ is also known
  \begin{equation}
      G_{\mathbb{R}^d}(x,x') = \frac{1}{2\pi}\qty(\frac{m}{2\pi r})^{\frac{d}{2}-1}K_{\frac{d}{2}-1}(mr),
  \end{equation}
  where $K_{\nu}(z)$ is the modified Bessel function of the second kind.
  For the real $r>0$ and $t>0$ case we can show that the previous identity holds
  \begin{equation}\label{eq:heat kernel Green's func}
      \int_0^\infty dt \; K_{\mathbb{R}^d}(x,x';t) = G_{\mathbb{R}^d}(x,x').
  \end{equation}
 In our double cone regularization, we should ensure that the identity (\ref{eq:heat kernel Green's func}) holds for $r$ such that $\mathrm{Re}\qty[r]>0$ and $\mathrm{Im}\qty[r]<0$. To achieve this, we consider deforming the contour of the integral away from the real axis. One possible contour satisfying these criteria is illustrated in Fig.\ref{fig:tcontour}.
This is justified as follows: as a mathematical fact, it is known that the Hankel function of the first kind $H_\nu^{(1)}(z)$ has an integral representation,
\begin{equation}
    \begin{split}
        H_\nu^{(1)}(z) = \frac{1}{i\pi}\int_{\Gamma'} \frac{dt'}{t'^{\nu+1}}e^{\frac{z}{2}\qty(t'-\frac{1}{t'})},
    \end{split}
\end{equation}
for $\mathrm{Re}\qty[z]>0$ \cite{Hankel_1995}. $\Gamma'$ is illustrated in Fig.\ref{fig:tcontour}. Let us set $z=imr$, where $\mathrm{Re}\qty[r]>0$ and $\mathrm{Im}\qty[r]<0$. Then we see $\mathrm{Re}\qty[imr]>0$ and $\mathrm{Im}\qty[imr]>0$. 
By combining these 
\begin{equation}
\begin{split}
    K_\nu(mr) &= \frac{i^\nu}{2}\int_{\Gamma'}\frac{dt'}{t'^{\nu+1}}e^{\frac{imr}{2}\qty(t'-\frac{1}{t'})}\\&= \frac{1}{2}\qty(\frac{r}{2m})^\nu \int_\Gamma \frac{dt}{t^{\nu+1}} e^{-m^2 t -\frac{r^2}{4t}}.\nn
\end{split}
\end{equation}
By setting $\nu=\frac{d}{2}-1$ and $t'= -\frac{2m^2}{imr} t$, we obtain (\ref{eq:heat kernel Green's func}) for $r$ such that $\mathrm{Re}\qty[r]>0$ and $\mathrm{Im}\qty[r]<0$ so that  $\Gamma$ is the correct contour for this region.\par
With this contour $\Gamma$, we revisit the computation above. As an important caveat, the $r$-integral
on page 3 with this contour $\Gamma$ faces an IR divergence since $\Gamma$ includes the Schwinger parameter with negative real $\Re[t]<0$. To this end, we need to introduce an IR cut-off $r_\infty$ for the integral. We take $r_\infty\to \infty$ after the $t$-integration. Having done the IR regularization, we obtain a finite $r$-integral for any $t$. Importantly, the resulting $t$-integral with this cut-off is finite in the $r_\infty\to \infty$ limit, and thus the entanglement entropy is IR finite.  To be explicit, we focus on the large $r$ and ignore the UV cut-off part for now, which is not relevant to the IR cutoff sensitivity of the entropy. Also, we consider the $\mathbb{Z}_2$ symmetry of the integral. Then, we compute the simplified integral
\begin{equation}
\int^{r_\infty} dr\; r\; e^{-\frac{r^2}{t}\sin^2{\frac{\pi k}{N}}} = \frac{t}{2\sin^2{\frac{\pi k}{N}}}e^{-\frac{{r_\infty}^2}{t}\sin^2{\frac{\pi k}{N}}}
\end{equation}

We can  see that this diverges for $r_\infty\to \infty$ with $\Re[t]<0$. However, we claim that after the $t$-integral, the $S_\mathcal{A}$ is finite even if we take $r_\infty \to \infty$. Indeed, we find
\begin{equation*}
\begin{split}
   &\quad \int_\Gamma \frac{dt}{t} \frac{1}{(4\pi t)^d} \frac{t}{2\sin^2 {\frac{\pi k}{N}}} e^{-\frac{{r_\infty}^2}{t}\sin^2 {\frac{\pi k}{N}}} e^{-tm^2}\\
   &= \frac{1}{(4\pi)^d\sin^2 {\frac{\pi k}{N}}} \qty(\frac{m}{r_{\infty}\sin{\frac{\pi k}{N}}})^{d-1} \, K_{d-1}\qty(mr_{\infty}\sin {\frac{\pi k}{N}})\\
    &\sim \qty({r_\infty})^{-d+\frac{1}{2}} e^{-mr_{\infty}\sin {\frac{\pi k}{N}}}\to 0, \quad \text{as $r_\infty\to \infty$.}
\end{split}
\end{equation*}
Here, we use the integral representation of the modified Bessel function $K_\nu(z)$ written above. For large $r_\infty$, it is known that 
\begin{equation}
    K_{\nu}(z) \sim \qty(\frac{\pi}{2 z})^{\frac{1}{2}}e^{-z}\qty(1+\order{\frac{1}{z}}).
\end{equation}

%%%%%%%%%%%%%%%%%%%%%%%%%%%%%%%%%%%%%%
\par Finally, we move to the evaluation of the entanglement entropy. By choosing contour $\Gamma$ in  Fig.\ref{fig:tcontour}, we obtain the finite expression for the R\'{e}nyi entropy,
\begin{align}\label{eq:ZN partition1}
&Z[\dcN]-\frac{1}{N} Z[\dc]\nn \\
&= \frac{i\pi^2 V_{d-2}}{(4\pi)^{\frac{d}{2}} N} \sum_{k=1}^{N-1}\frac{1}{\sin^{2}{\frac{k \pi}{N}}}\qty(-\frac{m}{\ep \sin{\frac{k \pi}{N}}})^{^{\frac{d}{2}-1}} \!H_{^{\frac{d}{2}-1}}^{(1)}\qty(2\ep m\sin{\frac{k\pi}{N}}).\nn\end{align}
As with the case for the momentum cut-off, we can explore the $\ep$ expansion. Let's focus on the scenario when $d$ is even. In this case, the $n$-th R\'{e}nyi entropy can be expanded as follows:
\begin{align}
S_{\mathcal{A}}^{(n)}&= \frac{\qty(\frac{d}{2}-1)!}{2}(-1)^{\frac{d}{2}-1}\frac{V_{d-2}}{\ep^{d-2}}\frac{1}{N}\sum_{k=1}^{N-1}\frac{1}{(\sin{\frac{k\pi}{N}})^{d-2}}+\cdots\nn\\
    %&+\cdots\\
    &+\frac{n+1}{3n}\frac{2\pi}{(4\pi)^\frac{d}{2}}\frac{(-1)^{\frac{d}{2}-1}}{\qty(\frac{d}{2}-1)!}\cdot V_{d-2}m^{d-2}\log\qty(\frac{1}{m\ep})\nn\\
    &+\frac{n+1}{3n}\!\frac{2\pi}{(4\pi)^\frac{d}{2}}\frac{(-1)^{\frac{d}{2}-1}}{\qty(\frac{d}{2}-1)!}\qty(\!\psi\qty(\!\frac{d}{2}-1)\!-\gamma+\frac{i\pi}{2}\!)V_{d-2}m^{d-2}\\&+\cdots.\nn
\end{align}
It's noteworthy that all terms can be divided by $N-1$, allowing us to safely define the entanglement entropy. When $d$ is odd, we see:
\begin{equation}
\begin{split}
S_{\mathcal{A}}^{(n)} &= \frac{2\pi}{(4\pi)^{\frac{d}{2}}}\frac{1}{N}\sum_{k=1}^{N-1}\frac{1}{\qty(\sin{\frac{k\pi}{N}})^d}\frac{(-1)^{\frac{d}{2}-1}\Gamma\qty(\frac{d}{2}-1)}{2}\frac{V_{d-2}}{\ep^{d-2}}\\
%&+\cdots\\
&+\cdots+\frac{n+1}{3n}\frac{\pi}{(4\pi)^{\frac{d}{2}}}\Gamma\qty(\frac{2-d}{2})\cdot V_{d-2}m^{d-2}+\cdots.
\end{split}\nn
\end{equation}
It is important to note that the divergent terms for odd $d$ are purely imaginary.

One crucial observation is that if divided by two (extract a half cone), the universal term of the R\'{e}nyi entropy obtained via our double cone regularization matches the one from the momentum cut-off (\ref{eq:momentum RE universal}).
%  \subsection{Imaginary Part of the Renyi entropy}

 As is customary, the density matrix of the right Rindler wedge in Minkowski space is naively expressed as
\begin{equation}
    \rho = \frac{1}{Z(m^2,\beta)} e^{-\beta K},\quad Z(m^2,\beta)= \Tr\qty[e^{-\beta K}],
\end{equation}
where $K$ represents a boost operator \cite{PhysRevD.14.870}. 
In the  double cone regularization, it is natural to replace the boost operator $K$ with the modified boost operator $\widetilde{K}$ \cite{Chen:2023hra}, which is defined as the translation of $\theta$ in the deformed metric \eqref{eq:our_double_cone},\eqref{eq:our_double_cone_ZN}
, yielding
\begin{equation}
    \rho = \frac{1}{\widetilde{Z}(m^2,\beta)} e^{-2\pi \widetilde{K}},\quad \widetilde{Z}(m^2,\beta)= \Tr\qty[e^{-\beta \widetilde{K}}] .
\end{equation}
It is important to note that if this holds true, $\rho$ is not a Hermitian operator since $\widetilde{K}$ is non-Hermitian. The eigenvalues of the $\widetilde{K}$ are called quasi-normal modes (here correspondingly they are complex valued). In Appendix A, we derive the spectrum of these quasi-normal modes and show that it is independent of our regularization parameter $\epsilon>0$. These quasi-normal modes determine the poles of the partition function $\widetilde{Z}$ \cite{Denef:2009kn}, leading to the formula
\begin{equation}
    \widetilde{Z}(m^2,\beta) = e^{\mathrm{Pol}(m^2)} \prod_{l\geq 1,\Vec{k}} \Gamma\left(1 + \beta\frac{i2r_\infty\xi + 2\left(l-\frac{1}{4}\right)}{2\pi }\right),\nn
\end{equation}
where $r_\infty$ is an IR cut-off, $\Vec{k}$ are momentum for the $\mathbb{R}^{d-2}$ directions and $\xi=\sqrt{\Vec{k}^2+m^2}$. $\beta$ is $2\pi$ or $2\pi n$ depending on assuming one-sheet or $n$-sheet geometry, respectively.
Here, we exclusively consider the quasi-normal modes ($l\geq1$) as $\mathrm{Im}[\widetilde{K}]<0$ as discussed in \cite{Chen:2023hra}. Despite the absence of the UV parameter $\epsilon$ in the quasi-normal modes, the analytic function $\mathrm{Pol}(m^2)$ includes the UV parameter. 
This discussion on non-Hermitian density matrices assures us that the Renyi entropy is not obliged to be a real value. Specifically, taking the limit $n\to 1$ yields a complex-valued entanglement entropy. However, in our scenario, entropic inequalities like subadditivity or strong subadditivity are not naively applicable due to the non-Hermitian nature of the density matrix. This is similar to a case that von Neumann entropy defined through transition matrix, called pseudo entropy, violates entropic inequalities \cite{Nakata:2020luh}.\vspace{1 pt}

\noindent 
 \emph{3.Double cone in  two-dimensional CFTs. }
 In two-dimensional conformal field theories, the entanglement entropy in the single interval case is universal 
  \be
S_A=\frac{c}{3}\log\frac{L}{\ep},\label{uni}
  \ee
 where the  $c$ is the central charge of the CFT and $L$ is the interval length. As elucidated in the preceding section, our computations in the free boson theory in two dimensions are consistent with the behavior expressed in (\ref{uni}), with $c=1$.

Several methodologies exist for deriving the universal behavior articulated in (\ref{uni}). Authors have employed techniques such as twist operators, as documented in \cite{Calabrese:2004eu}, and boundary states, allowing for  understanding from the algebraic structure of the Hilbert space, as explained in \cite{Ohmori:2014eia}.

\begin{figure}[h]
\begin{center}
\includegraphics[width=65mm]{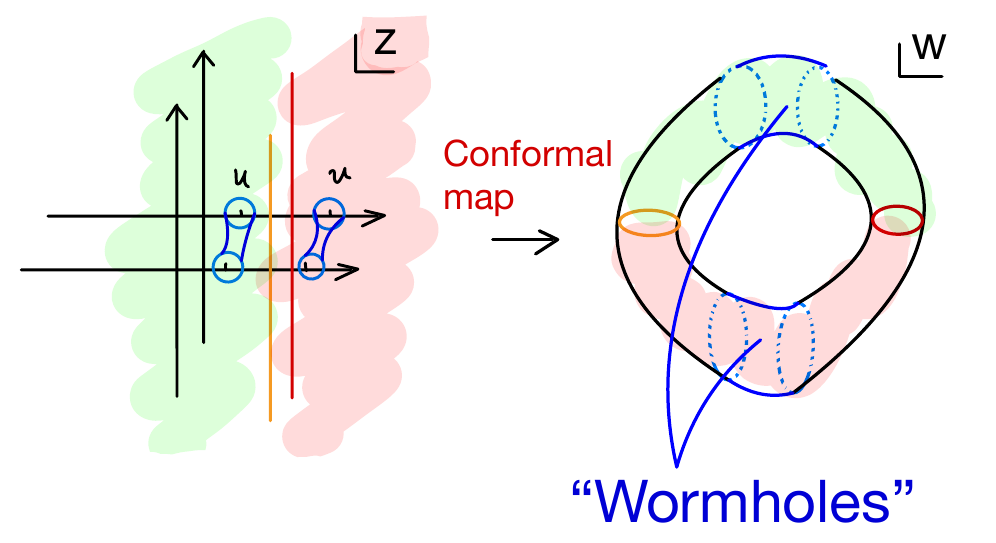}
\caption{Construction of the "torus" geometry connected by  "wormholes" via the  $i\ep$ prescription. The left figure shows the double cone geometry through "wormholes", which is made of two planes. Each planes describes $\Re[r]>0$ and $\Re[r]<0$ regions respectively. The right one describes the "torus" geometry connected by the "wormhole" throats. }
\label{gluing}
\end{center}
\end{figure}

Now, we intend to employ the $i\ep$ prescription in the replica manifold. We shall focus on the scenario of a flat two-dimensional plane, denoted as $\mathbb{R}^2$, hosting a unitary conformal field theory (CFT). It is noteworthy that our considerations encompass arbitrary two-dimensional CFTs, irrespective of their central charges. The metric of interest is expressed as follows:
\be
ds^2=dx^2+d\tau^2=dr^2+r^2d\theta^2.
\ee
We designate the subsystem $\mathcal{A}$ as $x\in[0,L]$ at $\tau=0$. To facilitate calculations, we introduce complex coordinates as follows:
\be
z=x+i\tau,\quad \Bar{z}=x-i\tau.
\ee
We begin by constructing the double cone geometry  resembling a "wormhole" with two throats attached around $z=0$ and $z=L$. In order to regulate the physics around the entangling surface (the boundary of the subsystem), we employ the $i\ep$ prescription in that region. The procedure is outlined as follows: We consider the radial coordinate around the endpoints of the interval, denoted as $z=0,L$ in Fig.\ref{gluing}. Similar to the flat plane case, we extend the radial direction into negative values using the $i\ep$ prescription
\be
z=(r-i\ep)e^{i\theta}.
\ee
One may wonder if this prescription is correct since we can also construct a double cone geometry in an $n$-sheet geometry. We can leverage a map
\be
\zeta=\left(\frac{z-L}{z}\right)^{\frac{1}{n}},\label{map}
\ee
to construct the $n$-sheet geometry.  Since we are in a flat space $\mathbb{R}^2$, we can create the double cone geometry using the $i\ep$ prescription, as done previously. In this scenario, the position of the "wormhole" throat slightly differs from $z=0$ and $z=L$ if we pull back to one-sheet geometry. For instance, employing a map (see (\ref{map})), we can construct a double cone geometry with the $i\ep$ prescription, which can then be pulled back to the one-sheet geometry. The position of the throat around $L$ is determined by:
\be
z=\frac{L}{1-(-i\ep e^{i\theta})^n},
\ee
and even in the $n\rightarrow1$ limit  this position differs from the position, where we originally introduced the $i\ep$ prescription
\be
z=L-i\ep e^{i\theta}.
\ee
Although the details of the throat positions may differ (and sometimes the shape of the throat may also differ in different regularization schemes), they do not change the leading terms in the entanglement entropy, as we will see below. The discussion below parallels the argument in \cite{Ohmori:2014eia}.

After introducing the double cone geometry with the $i\ep$ prescription for the one-sheet geometry, we can map it into the torus
\be
w=\log\frac{z}{L-z}.
\ee
Note that, since we are dealing with the complex manifold, the originally anti-holomorphic part is completely independent. Here and below, we consider the holomorphic part, but we find that the anti-holomorphic part gives the same contribution. After using this map, the length and circumference of the torus become $l=4\log\frac{L}{\ep}+\cdots$ and $2\pi$, respectively. The length of the torus varies depending on the details of the regularization described above, though it only affects the sub-leading terms in the $\frac{L}{\ep}$ expansion. We can also construct the $n$-sheet geometry and obtain a torus whose length and circumference are given by $l=4\log\frac{L}{\ep}+\cdots$ and $2\pi n$, respectively. Using a modular transformation, we can rescale the length and the circumference as $l=\frac{4}{n}\log\frac{L}{\ep}+\cdots$ and $2\pi$. We also comment that the sub-leading corrections to length can be imaginary valued in general depending on the detailed choice of the complex contour. It also depends on which manifold (one-sheet or $n$-sheet) we construct the double cone geometry. 

Then, the entanglement entropy reads
\be
S_{\mathcal{A}}=\lim_{n\to1}\frac{1}{1-n}\log\left(\frac{Z_n}{(Z_1)^n}\right),
\ee
 where $Z_n$ and $Z_1$ denote the partition function of the $n$ and one-sheet geometries. In the long length limit ($\log\frac{L}{\ep}\rightarrow\infty$), the partition function can be approximated by truncating the vacuum state propagation as discussed in \cite{Ohmori:2014eia,Cardy:2016fqc}\footnote{The universal term comes from the torus partition fucntion. See \cite{Cardy:2016fqc} for details.}
\be
S_{\mathcal{A}}=\lim_{n\to1}\frac{1}{1-n}\frac{c}{6}\frac{1-n^2}{n}\log\frac{L}{\ep}=\frac{c}{3}\log\frac{L}{\ep}.
\ee
We need to extract the half cone contribution, divide the result by two, and take into account the contribution from the anti-holomorphic part. Therefore, we obtain
\be
S_{\mathcal{A}}=\frac{c}{3}\log\frac{L}{\ep},
\ee
where we ignore the sub-leading corrections in the $\frac{L}{\ep}$ expansions, which are complex valued in general\footnote{In this 2-dim CFT case, the bulk and "boundary"(if we believe there is a corresponding boundary condition with the $i\ep$ prescription and divide the net value of the partition function by two) contributions decouple since we take the long torus length limit. We expect that this boundary entropy probably show monotonicity in some sence since in a holographic set-up this comes from the tension of de Sitter brane in the AdS space and the monotonicity follows from the energy condition of the AdS space. }.

\vspace{1 pt}
\noindent \emph{4.Conclusion.}
We presented an alternative derivation of the universal component of entanglement entropy using the $i\epsilon$ prescription applied to the Euclidean double cone geometry. Our calculations were demonstrated in the free boson theory in arbitrary dimensions and in two-dimensional CFTs. In the case of the free boson theory, we successfully reproduced the universal logarithmic and constant terms in even and odd dimensions, respectively. Notably, the constant terms in even dimensions were found to be imaginary. 

For two-dimensional CFTs, we derived the universal logarithmic term in the entanglement entropy. While entanglement entropy is typically expected to be real-valued, our results included imaginary terms. The interpretation of these imaginary terms requires further exploration,
%particularly in the context of von Neumann algebras, 
understanding the connection with complex geometries as well as their implications for the de Sitter holography.

Throughout this paper, our focus has been on elucidating universal behavior, and while we have encountered imaginary contributions in non-universal components, we anticipate that there may be a suitable method to define real-valued entanglement entropy even with the deformed complex contour. Additionally, applying our approach to different metrics and problems where careful control of the cut-off scale is necessary could provide further insights.\par
Also we may say that our complex deformation induce the complex modular operator which will be something new from the operator algebra perspective. That is, it may be formally possible that the type $\grethree$ von Neumann algebra of the subsystem of the QFT will be modified some algebra with complex modular operator, which is something new. 
\vspace{8 pt}
\begin{acknowledgments}
%%%%%%%%%%%%%%%%%%%%%%%%%%%%%
    \emph{Acknowledgments.---}We are grateful for Y.Taki and M.Watanabe in useful discussions at the early stage of this work and ongoing collaborations related to this work. We thank S.M.Ruan and T.Takayanagi for fruitful discussions. We also thank K.Doi, Z.Wei and ChatGPT 3.5 for revising our writings and  H.Liu for his insightful comments on the quasi-normal modes and non-Hermitian modular operators.
TK is supported by Grant-in-Aid for JSPS Fellows No. 23KJ1315. YS is supported by Grant-in-Aid for JSPS Fellows No.23KJ1337. 

%%%%%%%%%%%%%%%%%%%%%%%%%%%%%
\end{acknowledgments}
%%%%%%%%%%%%%%%%%%%%%%%%%%%%%

%%%%%%%%%%%%%%%%%%%%%%%%
%\bibliographystyle{apsrev4-2} 
\bibliographystyle{apsrev4-1}
\bibliography{PRL.bib}
%%%%%%%%%%%%%%%%%%%%%%%%
%%%%%%%%%%%%%%%%%%%%%%%%
%%%%%%%%%%%%%%%%%%%%%%%%

%%%%%%%%%%%%%%%%%%%%%%%%%%%%%%%%%%%%%%%%%%%%%%%%%%%%%%%%%%
%%%%%%%%%%%%%%%%%%%%%%%%%%%%%%%%%%%%%%%%%%%%%%%%%%%%%%%%%
%Appendix%%%%%%%%%%%%%%%%%
%%%%%%%%%%%%%%%%%%%%%%%%%%%%%%%%%%%%%%%%%%%%%%%%%%%%%%%
%%%%%%%%%%%%%%%%%%%%%%%%%%%%%%%%%%%%%%%%%%%%%%%%%%%%%%%%%%
%%%%%%%%%%%%%%%%%%%%%%%%%%%%%%%%%%%%%%%%%%%%%%%%%%%%%%%%%%
%\begin{center}
%	{\LARGE Supplemental Material}
%\end{center}
\pagebreak
\titlepage
\setcounter{page}{1}

%%%%%%%%%%%%%%%%%%%%%%%%%%%%%%%%%%%%%%%%%%%%%%%%%%%%%%%%
\section{Appendix A: Quasi-normal Modes on the Double Cone}\label{appA}

In this appendix, we explore the massive free field theory within the framework of a double cone geometry in Minkowski space. To see the complex spectrum of the modified boost operator $\Tilde{K}$, we consider to solve the wave equation on
\begin{equation}
     ds^2= dr^2+r^2 d\tau^2 +\sum_{i=2}^d (dx^i)^2 \in \gamma_{\mathrm{app}}
\end{equation}
with another complex contour $\gamma_{\mathrm{app}}$ in the FIG.\ref{fig:r-contour}. We expect the solution on the contour $\gamma$ which is used in the Renyi entropy computation is obtained by the analytic continuation from the solution on $\gamma_{\mathrm{app}}$. 
While our approach parallels that of \cite{Bah:2022uyz,Chen:2023hra}, we explicitly provide the exact solution. The scalar fields $\phi_l(r)$ and $\phi_r(r)$ are considered for $\mathrm{Re}\qty[r]>0$ and $\mathrm{Re}\qty[r]<0$, respectively. The wave equation to be solved is given by
\begin{equation}
    \qty( \frac{1}{\sqrt{r^2}}\partial_r \sqrt{r^2}\partial_r +\frac{1}{r^2}\partial_{\tau}^2+\sum_{i=2}^d \partial_i^2 -m^2)\phi=0,\; r\in \gamma_{\mathrm{app}}.
\end{equation}

Assuming physical $\mathbb{Z}_2$ symmetry between left and right, we propose the following ansatz for the equations:
\begin{equation}\label{eq:junction}
     \phi_r(r)= e^{-\omega\tau}e^{i\Vec{k}\cdot \Vec{x}}F_\omega(r),\quad \text{for $r>0$},\quad \phi_l(r)= \Lambda(\omega)e^{-\omega\tau}e^{i\Vec{k}\cdot \Vec{x}}F_\omega(\abs{r}),\quad \text{for $r<0$},
\end{equation}
where $r$ is a real variable, $\Vec{k},\Vec{x}$ denote the momentum and coordinate for the vertical direction and $\xi=\sqrt{\Vec{k}^2+m^2}$. The solutions for the wave equation take the form of a linear combination of Bessel functions:
\begin{equation}
    F_\omega(r)= A(\omega) J_{i\omega}(-ir\xi)+B(\omega)Y_{i\omega}(-ir\xi).
\end{equation}

To ensure the analytic continuation of the right field to the lower half-plane matches the left field, as expressed in Eq.~\eqref{eq:junction}, we impose the conditions:
\begin{equation}
 \begin{split}
     &J_{i\omega}(e^{-i\pi}R)= e^{\pi\omega}J_{i\omega}(R),\\
     &Y_{i\omega}(e^{-i\pi}R)= e^{-\pi\omega}Y_{i\omega}(R)-2i\cosh{\pi\omega}J_{i\omega}(R).
 \end{split}
\end{equation}
For the case where $i\omega\notin\mathbb{Z}$, this leads to
\begin{equation}
 \begin{split}
     \phi_l(e^{-i\pi}r)&= e^{-i\omega t}\qty(\qty(e^{\pi\omega}A(\omega)-2i\cosh{\pi\omega}B(\omega))J_{i\omega}(-ir\xi)+e^{-\pi\omega}B(\omega)Y_{i\omega}(-ir\xi)),\\
     \phi_r(-r) &= e^{-i\omega t}\Lambda(\omega)\qty( A(\omega) J_{i\omega}(-ir\xi)+B(\omega)Y_{i\omega}(-ir\xi)).
 \end{split}
 \end{equation}

Given the independence of the two Bessel functions for all orders, we deduce the relations:
\begin{equation}
     \begin{split}
e^{\pi\omega}A(\omega)-2i\cosh{\pi\omega}B(\omega) &= \Lambda(\omega) A(\omega),\\
e^{-\pi\omega}B(\omega) &= \Lambda(\omega) B(\omega).
     \end{split}
 \end{equation}

Suppose $B(\omega)\neq 0$, leading to $\Lambda(\omega)=e^{-\pi\omega}$. The first equation becomes
\begin{equation}\label{eq:AB}
     \sinh{\pi\omega} A(\omega)-i\cosh{\pi\omega}B(\omega)=0.
 \end{equation}

Introducing the IR cutoff $r=r_\infty$ and demanding the Dirichlet boundary condition $\phi(r=r_\infty)=0$ yields the solution:
\begin{equation}
F_\omega(r)= c(\omega)\qty(Y_{i\omega}(-ir_\infty\xi)J_{i\omega}(-ir\xi)-J_{i\omega}(-ir_\infty\xi)Y_{i\omega}(-ir\xi)).
\end{equation}

We find $B(\omega)=-\frac{J_{i\omega}(-ir_\infty\xi)}{Y_{i\omega}(-ir_\infty\xi)}A(\omega)$. Substituting this into Eq.~\eqref{eq:AB}, we obtain
\begin{equation}
\sinh{\pi\omega}Y_{i\omega}(-ir_\infty\xi)+i\cosh{\pi\omega}J_{i\omega}(-ir_\infty\xi)=0.
\end{equation}

For the case where the IR cutoff is very large $r_\infty \xi \gg 1$, and using the asymptotic form of the Bessel functions
\begin{equation}
    \begin{split}
        J_{i\omega}(z)&\sim \sqrt{\frac{1}{\pi z}}\cos{\qty(z-\frac{i\omega\pi}{2}-\frac{\pi}{4})}+\cdots,\\
        Y_{i\omega}(z)&\sim \sqrt{\frac{1}{\pi z}}\sin{\qty(z-\frac{i\omega\pi}{2}-\frac{\pi}{4})}+\cdots,
    \end{split}
\end{equation}
we obtain
\begin{equation}
\sin{i\pi\omega}\sin{\qty(-ir_\infty\xi-\frac{i\omega \pi}{2}-\frac{\pi}{4})}-\cos{i\pi\omega}\cos{\qty(-ir_\infty\xi-\frac{i\omega \pi}{2}-\frac{\pi}{4})}=0.
\end{equation}

This implies
\begin{equation}
\begin{split}
    i\frac{\omega \pi}{2}-&i r_\infty \xi -\frac{\pi}{4} =\qty(l-\frac{1}{2})\pi,\quad l\in\mathbb{Z},\\
    \omega_l &= \frac{2 r_{\infty} \xi}{\pi}-2i\qty(l-\frac{1}{4}),\quad l\in\mathbb{Z}.
\end{split}
\end{equation}
Thus, we have obtained the quasi-normal mode. In the case where $i\omega\in\mathbb{Z}$, the analytic continuation of the Bessel functions of the second kind is slightly modified:
\begin{equation}
    Y_l(e^{-i\pi}R)=(-1)^l\qty(Y_l(R)-2i J_l(R)).
\end{equation}
From the junction condition \eqref{eq:junction}, we deduce
\begin{equation}
\begin{split}
    &(-1)^l\qty(A_l-2i B_l)= \Lambda_l A_l,\\
    &(-1)^l B_l =\Lambda_l B_l.
\end{split}
\end{equation}
Considering the fall-off condition $B_l=-\frac{J_l(-ir_\infty \xi)}{Y_l(-ir_\infty \xi)}A_l$ and aiming for nontrivial solutions with $A_l\neq0,B_l\neq 0$, we find $\Lambda_l =(-1)^l$. However, this implies $B_l=0$, and consequently, we do not have a non-trivial solution in this case.
\end{document}